\documentclass[toc]{PoS}

\usepackage{tikz}

\usepackage[matrix,arrow]{xy}

\usepackage{epsfig}

\usepackage{amssymb,amsthm}
\usepackage{amsmath}
\usepackage{amscd}
\usepackage{amsfonts}
\usepackage{latexsym}
\usepackage{graphics}
\usepackage{color}
\usepackage{bbm}
\usepackage{bm}

\usepackage{xypic}
\usepackage{mathrsfs}

\input{xy}
\xyoption{all}

\def\={\ =\ }
\def\dd{{\rm d}}

\def\e{{\,\rm e}\,}
\newcommand{\mbf}[1]{{\boldsymbol {#1} }}

\DeclareMathOperator{\RH}{\rm H}

\def\ii{{\,{\rm i}\,}}

\newcommand{\bbr}{\mathbb{R}}

\newcommand{\calD}{\mathcal{D}}

\newcommand{\call}{\mathcal{L}}

\newcommand{\calo}{\mathcal{O}}
\newcommand{\calm}{\mathfrak{M}}

\newcommand{\calq}{\mathcal{Q}}
\newcommand{\calh}{\mathcal{H}}

\newcommand{\calT}{\mathcal{T}}

\newcommand{\frg}{\mathfrak{g}}
\newcommand{\frh}{\mathfrak{h}}

\newcommand{\CF}{\mathcal{F}}

\newcommand{\CCG}{\mathscr{G}}

\newcommand{\CL}{\mathcal{L}}
\newcommand{\CM}{\mathcal{M}}

\newcommand{\IZ}{\mathbb{Z}}

\def\frh{{\mathfrak{h}}}

\newcommand{\bbz}{{\mathbb Z}}

\def\e{\epsilon}

\def\beq{\begin{equation}}
\def\eeq{\end{equation}}
\def\bea{\begin{eqnarray}}
\def\eea{\end{eqnarray}}
\def\beqa{\begin{eqnarray*}}
\def\eeqa{\end{eqnarray*}}

\renewcommand{\time}{{\times}}
\renewcommand{\e}{\,\mathrm{e}\,}
\newcommand{\im}{\,\mathrm{i}\,}

\newcommand{\delder}[1]{\frac{\delta}{\delta #1}}

\newcommand{\intps}{\int_{{\cal M}} \, \dd^{2d}x \ }

\def\>{\rangle}
\def\<{\langle}
\def\+{\dagger}
\def\={\ =\ }

\title{Nonassociative geometry and twist deformations in
  non-geometric string theory}

\ShortTitle{Nonassociative geometry in non-geometric string theory
  \hspace{1.25in} {\rm Mylonas, Schupp \& Szabo}}

\author{Dionysios Mylonas, \
  Peter Schupp \ and \ Richard J. Szabo\thanks{Speaker.}%
        \\ Heriot-Watt University, Edinburgh, U.K. \\
        E-mail: \email{dm281@hw.ac.uk , R.J.Szabo@hw.ac.uk}\\
Jacobs University
  Bremen, Germany\\
Email: \email{p.schupp@jacobs-university.de}}

\abstract{
We describe nonassociative deformations of geometry probed by closed strings in non-geometric flux
compactifications of string theory. We show that these non-geometric backgrounds can be geometrised through the
dynamics of open membranes whose boundaries propagate in the phase space of the target
space compactification, equiped with a twisted Poisson structure. The
effective membrane target space is determined by the standard Courant
algebroid over the target space twisted by an abelian gerbe in
momentum space. Quantization of the
membrane sigma-model leads to a proper quantization of the non-geometric background, which we relate to
Kontsevich's formalism of global deformation quantization that
constructs a noncommutative nonassociative star product on phase space. We construct Seiberg--Witten type maps between
associative and nonassociative backgrounds, and show how they may realise a nonassociative deformation of
gravity. We also explain how this approach is related to the
quantization of certain Lie 2-algebras canonically associated to the twisted Courant
algebroid, and cochain
twist quantization using suitable quasi-Hopf algebras of symmetries in the
phase space description of $R$-space which constructs a Drinfel'd
twist with non-trivial 3-cocycle. We illustrate and apply our formalism
to present a consistent phase space formulation of nonassociative quantum mechanics.\\ \\
{\tt Report number: \ EMPG--14--5}
}

\FullConference{Third International Satellite Conference on Mathematical Methods in Physics -- ICMP 2013\\
		21--26 October, 2013\\
		Londrina, PR, Brazil}

\begin{document}

\section{String geometry\label{Intro}}

Strings see geometry in different ways than point particles
do because of their extended nature. This feature has led to many novel examples of \emph{string geometry} which
involve modifications, sometimes radical, of standard geometric
structures. One of the first examples of such string symmetries was
T-duality, which due to string winding modes implies
that large and small compact directions are indistinguishable in string theory. Together with related phenomena such as mirror symmetry,
these predictions from string theory have opened up many unexpected
developments in geometry. In many such cases string theory is usually studied
in regimes whereby a geometric description is available. However, via
T-duality string theory also admits
non-geometric backgrounds as consistent solutions~\cite{Dabholkar2006}.

It has been
long hoped that closed strings, and also other degrees of freedom in
string theory, provide good probes of Planck scale quantum geometry
where the classical notions of general relativity break down. In fact,
a simple argument for the semi-classical quantization of gravity suggests that
spacetime coordinates themselves should be subjected to an uncertainty
principle $\Delta x  \geq  \ell_P$; here $\ell_P$ is some
  suitable fundamental length scale such as the Planck scale or the string
length. This requirement ensures that one cannot localize events in
spacetime so as to produce strong gravitational fields that hide the
events to distant observers. In this article we shall review how such spacetime
uncertainties are related to noncommutative spacetime
structures and to non-geometry in string theory. We describe how
a complementary target space approach to quantum gravity based on
noncommutative geometry is related to certain non-geometric
generalizations of the target spacetime in string theory.

A precise realisation of noncommutative geometry in string theory
first appeared in the dynamics of the \emph{open} string sector. It was realised
some time ago that D-branes in constant background Neveu--Schwarz $B$-fields
  provide concrete dynamical realisations of noncommutative
  spaces~\cite{Chu1999,Schomerus1999,Seiberg1999}. The low-energy
  dynamics of this system is described
  by a noncommutative gauge theory on the D-brane worldvolume, together with Seiberg--Witten maps
  which relate them to ordinary (commutative) gauge theories. These
  developments have led to a flurry of investigation over the past 15
  years into the structures and properties of these noncommutative
  field theories; see e.g.~\cite{Douglas2001,Szabo2003} for early
  pedagogical reviews on the subject. The key feature of these
  realisations is that they involve a low-energy regime of string theory which
  decouples all massive open string states and all closed string
  modes while still retaining the
  effects of noncommutativity. The
  emergent noncommutative geometry in this limit is completely
  analogous to that of electron coordinates in a constant magnetic
  background in the lowest Landau level (see e.g.~\cite{Szabo2004}).

Analogous structures can also arise in the dynamics of the \emph{closed} string sector as a
result of the intertwining of momentum and winding modes in
non-geometric backgrounds. To understand this point, let us recall how noncommutativity in the open string sector arises
following~\cite{Seiberg1999}. In the Seiberg--Witten scaling limit which
decouples the bulk closed string modes from the open string boundary degrees of
freedom, the two-point function for string field insertions on the boundary
of a disk in tree-level open string perturbation theory depends only
on the relative ordering of the insertion points, and not on their
actual positions. This means that the two-point function is a
well-defined target space entity that is independent of the worldsheet
coordinates. It leads to a noncommutative 2-bracket between string
coordinates, with the deformation provided by a bivector which is
determined by the inverse of the constant two-form $B$-field. Quantization of
this 2-bracket yields a star product of fields on D-brane worldvolumes and
noncommutative gauge theory as a deformation of the low-energy effective field theory in the open string sector.

The natural question of how analogous structures can arise in the
closed string sector has remained somewhat unclear until the recent
works~\cite{Blumenhagen2010,Lust2010,Blumenhagen2011}. At tree-level
in closed string perturbation theory, the two-point function on a
sphere depends explicitly on the worldsheet coordinates in the
low-energy limit, and so is not
a bonafide target space quantity. On the other hand, the three-point
function depends only on the relative orientation of the three
insertion points on the sphere, and hence a 3-bracket structure
emerges on target space. It is natural to identify this 3-bracket as a
measure of nonassociativity of the closed string coordinates, and in this
case the deformation is provided by a trivector induced by a
non-geometric $R$-flux which is T-dual to the constant three-form
$H$-flux of a Neveu--Schwarz $B$-field. Quantization of this 3-bracket
structure is anticipated to lead to closed string nonassociative
gravity as a deformation of the low-energy effective field theory in
the closed string sector.

In contrast to open string noncommutativity, which can arise in
on-shell string scattering amplitudes because only the cyclic ordering
of vertex operators is conformally invariant,
the nonassociativity observed here is a truly off-shell phenomenon. Once
momentum conservation in tachyon scattering amplitudes is taken into
account, all traces of nonassociativity disappear and the usual
crossing symmetry of correlation functions in two-dimensional
conformal field theory is recovered. The appearence of
nonassociativity should be regarded as a
feature whose consistency induces constraints, such as flux
quantization, and teaches us something about the nature of
non-geometric string theory. In the following we shall tackle
this nonassociative deformation of geometry head on and describe
techniques for describing its proper quantization, which should be
regarded as a type of short distance spacetime quantization which is
not seen by strings due to their finite intrinsic length and
resolution; in this sense these non-geometric quantum geometries can
arise as consistent string backgrounds. In particular, we
shall show how nonassociativity can be accounted for via the
derivation of \emph{dynamical} star-products, via a categorified version
of Weyl quantization, and through quasi-Hopf cochain twist
quantization techniques. Our quantization obeys the requisite
cyclicity requirements of two-dimensional conformal field theory, and
can be used to formulate a consistent nonassociative version of
quantum mechanics. We will describe several analogues of the
Seiberg--Witten map interpolating between associative and
nonassociative theories, particularly in the framework of a
nonassociative theory of gravity, and in the context of extending
these structures to non-constant backgrounds.

The appearence of nonassociative geometry in string theory and of systematic
noncommutative deformations of gravity is not new. Nonassociative
gauge theories arise naturally in the \emph{open} string sector when
D-branes are placed in a non-constant $B$-field background; the
nonassociative deformation in this case is controlled by the three-form
$H$-flux $H=\dd B\neq0$ through Kontsevich's deformation quantization
of $H$-twisted Poisson
structures which is reproduced by the correlation functions of open
string tachyon vertex operators~\cite{Cornalba2002,Ho2001,Herbst2001}. These open string
nonassociative spaces are described locally by associative algebras
and noncommutative gerbes~\cite{Aschieri2010} with characteristic class
$[H]$ via a patching of the twisted Poisson manifold. Cyclicity
appears and
nonassociativity disappears in on-shell tachyon scattering amplitudes
by using the Dirac--Born--Infeld field equations on the D-brane~\cite{Herbst2001,Herbst2004}. On the other hand,
noncommutative gravity has been formulated within the cocycle twist
deformation framework of~\cite{Aschieri2005} which gives a systematic
formulation of differential geometry and general relativity on
noncommutative spacetime. However, it is argued
by~\cite{AlvarezGaume2006} that the twisted diffeomorphisms
underlying the symmetries of these associative noncommutative
deformations do not arise as physical symmetries of string
theory, and that the low-energy effective field theory contains terms which
cannot be accounted for by a noncommutative deformation of gravity. Although all of these structures will
appear in the following, as long as we are looking for
(nonassociative) quantizations of gravity within the realm of string
theory, the mechanisms and features of the theory we develop in the
following will be rather different in nature.

\section{Non-geometric string backgrounds\label{sec:NGFB}}

\subsection{Non-geometric flux compactification}

Compactifications of string theory are required to relate
them to observable phenomenology and cosmology. Flux compactifications
include $p$-form fluxes along the compact directions; they stabilize
moduli and can lead to generalized topological as well as geometric structures wherein open
neighbourhoods are patched together by string symmetries. In
particular, by demanding T-duality covariance, Neveu--Schwarz
$H$-fluxes naturally lead to non-geometric fluxes. While the
standard geometric flux compactifications do not give all four-dimensional
supergravity theories, non-geometric fluxes enable more gaugings and
avoid many no-go theorems. In this section we describe the emergence
of non-geometry in flux compactifications of string theory, and how
non-geometric $Q$-fluxes lead to closed string noncommutativity.

The prototypical string background is provided by the three-torus $T^3$ with
constant $H$-flux and dilaton field, which by application of the
B\"uscher rules gives rise to geometric and non-geometric fluxes via
the T-duality chain~\cite{Hull2005,Shelton2005}
\beq
H_{ijk} \ \xrightarrow{ \ T_i \ } \ f^i{}_{jk} \ \xrightarrow{ \ T_j \
} \ Q^{ij}{}_k \ \xrightarrow{ \ T_k \ } \ R^{ijk} \ .
\label{eq:Tdualitychain}\eeq
Here $T_i$, $i=1,2,3$ denotes a T-duality transformation along the
$i$-th cycle of $T^3$, which in each step maps the flux to a new flux
with a raised index; geometrically this means that a given differential form component is
dualised to a vector field component. Let us run through the geometric
and non-geometric interpretations of each duality frame in the chain
(\ref{eq:Tdualitychain}).

The first member of the T-duality chain is the original flat torus $T^3$
with \emph{$H$-flux} $H=\dd B$. Abelian fluxes in string theory obey
analogues of the Dirac quantization condition (see
e.g.~\cite{Szabo2012}), and hence the three-form determines a
cohomology class $[H]\in\RH^3(T^3;\IZ)=\IZ$ which is the
characteristic class of a gerbe. Such a gerbe lies within the realm of
what we shall consider as a geometric background.

The next member of the chain involves a \emph{metric flux} $f$, which
determines a torsion in the geometry through the Cartan--Maurer equations
\beqa
\dd e^i= -\mbox{$\frac12$} \, f^i{}_{jk}\, e^j\wedge
  e^k
\eeqa
for one-forms $e^i$ dual to a local frame of vector fields $e_i$ with
$[e_i,e_j]=f^k{}_{ij}\, e_k $. In this T-duality frame the $B$-field
vanishes and the geometry is that of a twisted torus or Heisenberg
nilmanifold, which is a circle bundle of degree $[H]$ over a two-torus
$T^2$~\cite{Scherk1979,Kachru2003}. Hence this is still a geometric frame.

The situation becomes more interesting at the next member of
(\ref{eq:Tdualitychain}) which involves non-geometric
\emph{$Q$-flux}; closed strings in $Q$-space can be locally modeled as
a $T^2$ bundle over a circle $S^1$. In this case the closed string momentum and winding
  modes become entangled, and the background is called a \emph{T-fold}
  as the transition functions between local charts now involve stringy T-duality
  transformations~\cite{Hellerman2004,Dabholkar2003,Hull2005}. In this instance the metric and $B$-field are
  well-defined locally but not globally; in particular, the
  $T^2$-fibre is glued back to itself by an ${\rm SL}(2,\IZ)$ transformation
  as one winds around the base $S^1$. In~\cite{Mathai2004,Grange2007,Brodzki2009} it is argued
  that this non-geometry at the topological level can be regarded
  globally as a
  fibration over $S^1$ by
  noncommutative two-tori $T_\theta^2$ with the commutation
  relations $[x^i,x^j]= \ii Q^{ij}{}_k\, x^k$ among local coordinates.
  Here $\theta^{ij}(x)= Q^{ij}{}_k\, x^k$ determine a field of local
  noncommutativity parameters parametrized by the coordinates of the
  base $S^1$; this bivector is naturally dual to the $B$-field which
  is a potential for the original $H$-flux in the T-duality
  chain. These topological arguments are based in the open string
  sector in the sense that they exploit the realisation of T-duality
  in the twisted K-theory of certain $C^*$-algebras; in particular, in
  the classical limit $\theta=0$ they realise a geometric space only
  up to Morita equivalence. Later on we shall
  see how to regard this identification more concretely and physically
  in the context
  of closed strings which wind in the non-geometric background.

The final member of the T-duality chain (\ref{eq:Tdualitychain})
involves the non-geometric \emph{$R$-flux} and leads to a background
which is not even locally geometric~\cite{Shelton2005}. One may object
to the existence of this background because a $B$-field which sources a
constant non-zero $H$-flux is non-constant, hence the final direction
is not a Killing isometry of the background and the standard B\"uscher
rules cannot be applied to the $Q$-flux background. However, the
prescription is completely well-defined at the level of worldsheet
conformal field theory, and the final T-duality transformation is
performed by flipping the sign of the corresponding right-moving
closed string coordinate; the failure of the B\"uscher rules simply
reflects the absence of local geometric structures. It is argued
by~\cite{Bouwknegt2006,Ellwood2006} that the globalisation of this
non-geometry is a (topological) nonassociative three-torus, regarded
as a fibration over a point in which the local fibre
coordinates obey a 3-bracket relation of the form $[x^i,x^j,x^k]=
R^{ijk}$; in this setting nonassociativity is realised in terms of
twisted convolution products in a Busby--Smith convolution
$C^*$-algebra, which again reduces in the classical limit to a
geometric setting only up to Morita equivalence. Later on we shall find more concrete physical
realisations of such nonassociative spaces in the context of closed
strings which propagate in the non-geometric background. We
are only beginning to understand the (non-)geometry of $R$-flux
backgrounds in terms of new types of spatial noncommutative and
nonassociative structures.

\subsection{Magnetic backgrounds in quantum mechanics}

To help understand the meaning of a nonassociative geometry, we shall
now describe a simple magnetic field analog of nonassociativity in the
ordinary quantum mechanics of point particles. This
example illustrates that in fact nonassociativity arises naturally in
the wild, and how it should be interpreted in the context of closed
string non-geometric backgrounds. For this, let us start by recalling
that the noncommutative deformations of D-brane worldvolumes induced
by open strings in constant $B$-field backgrounds in the
Seiberg--Witten scaling limit have a natural analog in the physics of
the Landau problem (see e.g.~\cite{Szabo2004}). For charged particles
constrained to move in two dimensions under the influence of a
perpendicularly applied constant magnetic field ${\mbf B}=
B\,\hat{z}$, the strong field limit
projects the system to the lowest Landau level and the effective
Lagrangian is of first order in time derivatives of the particle
coordinates $(x,y)$; canonical quantization thereby gives a noncommutative
position space with relations $[x,y]= \im\theta$, where the
noncommutativity parameter $\theta$ is proportional to $B^{-1}$.

A more general magnetic field analog was proposed some time ago by
Jackiw~\cite{Jackiw1985}. Charged particles in three dimensions experience a magnetic field $\mbf B$ (with
  sources) via the Lorentz force law $\dot{\mbf \pi}\= \frac e{m} \, {\mbf
    \pi}\times {\mbf B}$ for the \emph{physical} (gauge covariant)
  momentum $\mbf\pi$. The Hamiltonian $\mathcal{H}= \frac1{2m}\, {\mbf \pi}^2$
  generates the Lorentz force $\dot{\mbf\pi}=\ii
  [\mathcal{H},{\mbf\pi}]$ only for a \emph{noncommutative momentum
    space} described by the phase space commutation relations
\beq
[x^i,x^j]=0 \quad , \quad [x^i,\pi_j]=\im \hbar \, \delta^i{}_j
\quad , \quad [\pi_i,\pi_j] = \ii \hbar\, e\,
\epsilon_{ijk}\, B^k \ .
\label{eq:ncmomspace}\eeq
In a magnetic background translation invariance is lost, but there is
still a gauge symmetry under the magnetic translations $U({\mbf
  a})=\e^{\frac\ii\hbar\, {\mbf a}\cdot{\mbf
    \pi}}$. It follows from the commutation relations and the
Baker--Campbell-Hausdorff formula that they do not
commute,
\beqa
U({\mbf a}_1)\, U({\mbf a}_2)= \e^{-\frac{\ii e}\hbar \, \Phi_{{\mbf
      a}_1,{\mbf a}_2}} \, U({\mbf
  a}_1+{\mbf a}_2) \ ,
\eeqa
where $\Phi_{{\mbf a}_1,{\mbf a}_2}= \frac12\, ({\mbf a}_1\times {\mbf
  a}_2) \cdot {\mbf B}$ is the magnetic flux through the (infinitesimal) triangle $\langle{\mbf a}_1,{\mbf
  a}_2\rangle$ spanned by the two vectors. This is simply the well-known result that the wavefunctions
of a particle in a magnetic background only carry a projective
representation of the translation group; in particular, the projective
phase is a 2-cocycle of the abelian group of translations.

More dramatically, nonassociativity generically arises through the
violation of the Jacobi identity, as can be seen by computing the
Jacobiator
\begin{eqnarray*}
[\pi_i,\pi_j,\pi_k]:= [\pi_i,[\pi_j,\pi_k]] +
[\pi_j,[\pi_k,\pi_i]] + [\pi_k,[\pi_i,\pi_j]] = -e\, \hbar^2
\, \epsilon_{ijk}\,
\nabla\cdot{\mbf B} =: -e\, \hbar^2\, H_{ijk} \ .
\end{eqnarray*}
As a consequence, the magnetic translations do not associate and have
an associator given by
\beqa
\big(U({\mbf a}_1)\, U({\mbf a}_2) \big)\, U({\mbf a}_3 )=
\e^{-\frac{\ii e}\hbar \, \Phi_{{\mbf a}_1,{\mbf a}_2, {\mbf a}_3}}\, U({\mbf a}_1)\, \big(U({\mbf a}_2) \, U({\mbf
  a}_3) \big)
\eeqa
where $\Phi_{{\mbf a}_1, {\mbf a}_2, {\mbf a}_3}=\frac16\, \big(({\mbf
  a}_1\times {\mbf a}_2)\cdot {\mbf a}_3\big)\, \nabla\cdot{\mbf B}$
is the magnetic flux through the (infinitesimal) tetrahedron $\langle {\mbf a}_1,{\mbf
  a}_2, {\mbf a}_3\rangle$ spanned by the three vectors. Here the
projective phase is a 3-cocycle of the abelian group of translations.

There are now two cases to consider. If we demand that the magnetic field
satisfies Maxwell's equations, then {$\nabla\cdot{\mbf B}=0$}; in this
case there are no magnetic sources, no flux, and
  associativity persists. Then the physical momentum operators
  have the standard representation ${\mbf \pi}=-\ii \hbar\, \nabla-e\,
  {\mbf A}$ on ${\rm L}^2$-wavefunctions in terms of a vector
  potential $\mbf A$ for the magnetic field; in particular, for a constant magnetic field ${\mbf B}=
B\,\hat{z}$ the commutation relations (\ref{eq:ncmomspace}) in
the strong field limit reproduce the noncommuting coordinates of the
lowest Landau level~\cite{Szabo2004}. On the other hand, if
  {$\nabla\cdot{\mbf B} \neq 0$} then magnetic monopoles are present,
  and nonassociativity of the magnetic translations persists unless
\beqa
\mbox{$\frac e\hbar$} \, \Phi_{{\mbf a}_1,{\mbf a_2},{\mbf a}_3} \ \in
\ \pi\, \bbz \ .
\eeqa
This is simply the {Dirac quantization
  condition} for magnetic charge; in this context it ensures the
basic postulates of quantum mechanics, wherein associativity of
operators is required. In this instance associativity of the global
translations is satisfied, even though the infinitesimal translations
still violate the Jacobi identity.

We shall see in the following that the same sort of nonassociative
relations come up in the non-geometric flux models obtained via T-duality
from a constant $H$-flux background, though in a ``dual'' sense
through nonassocitivity of coordinate space rather than momentum space. The analogues of the Dirac
quantization condition for $R$-flux backgrounds are described in the context of Matrix theory
compactifications in~\cite{Chatzistavrakidis2013} and of double field
theory in~\cite{Blumenhagen2013}. On the other hand, a consistent
nonassociative
version of quantum mechanics based on the phase space quantization of
the $R$-flux string model can be developed following~\cite{Mylonas2013} and shown
rigorously to lead to minimal volume uncertainty relations $\Delta
x^i\, \Delta x^j\, \Delta x^k\geq \frac32\, \hbar^2 \, R^{ijk}$
characterizing a coarse-graining of the non-geometric $R$-flux
background; this is consistent with the fact that these backgrounds do
not allow the introduction of point-like objects~\cite{Ellwood2006}. The parallels between
nonassociative parabolic $R$-flux string vacua and the dynamics of charged particles
in uniform magnetic charge distributions is elucidated in~\cite{Bakas2013}.

\subsection{Geometry of $n$-algebras}

The deformations of geometry we have been describing involve certain
higher generalizations of algebraic structures, in particular certain
$n$-bracket structures. We shall now explain
how these extensions fit into concrete geometric frameworks, and then
describe a means to quantize them.

Let us begin by recalling that a {Nambu--Poisson structure} on a
smooth manifold $M$ is an $n$-Lie algebra structure $\{-,\dots,-\} :
  C^\infty(M)^{\wedge n} \rightarrow C^\infty(M)$, which means that it
  satisfies the {fundamental identity}
$$
\big\{f_1,\dots ,f_{n-1},\{g_1,\dots ,g_n\}\big\}= \big\{\{f_1,\dots
,f_{n-1},g_1\},\dots ,g_n \big\} + \dots +\big\{g_1,\dots
,\{f_1,\dots ,f_{n-1},g_n\} \big\} \ .
$$
In addition, it is required to obey the {generalized Leibniz rule}
\begin{eqnarray*}
\{f \,g,h_1,\dots ,h_{n-1}\}\=f\,\{g,h_1, \dots
 ,h_{n-1}\}+\{f,h_1, \dots ,h_{n-1}\} \, g \ .
\end{eqnarray*}
These properties imply that a Nambu--Poisson $n$-bracket is determined
via a Nambu--Poisson $n$-vector $\Pi=\frac1{n!}\, \Pi^{i_1\cdots
  i_n}(x)\, \partial_{i_1}\wedge \cdots\wedge \partial_{i_n}$ as
$$
\{f_1,\dots,f_n\}=\Pi(\dd f_1,\dots,\dd f_n)= \Pi^{i_1\cdots
  i_n}(x)\, \partial_{i_1}f_1 \cdots \partial_{i_n}f_n \ .
$$
In this paper our prominent example will be the flat backgrounds $M=
\bbr^3$ or $M=T^3$ equiped with the Nambu--Poisson
3-bracket~\cite{Nambu1973}, which is defined on coordinate functions
in terms of a constant trivector $R=\frac16\,
R^{ijk}\, \partial_i\wedge\partial_j \wedge\partial_k$
by
$$
\{x^i,x^j,x^k\}=  R^{ijk} \ ,
$$
and extended by linearity and the generalized Leibniz rule; its
quantization gives the \emph{Nambu--Heisenberg algebra}. For further
details about the quantization of generic Nambu--Poisson structures,
see e.g.~\cite{DeBellis2010} and references therein.

We will set up a framework involving a suitable generalization of
geometric quantization for these higher bracket structures; this
involves the notion of \emph{multisymplectic
  manifolds}. An $n$-plectic manifold is a manifold $M$ equiped with a
  closed $n+1$-form $\omega$ which obeys the nondegeneracy condition:
  $\omega(v,-) =0$ if and only if $v=0$. In this parlance a
  $1$-plectic manifold is the usual symplectic manifold, while a
  $2$-plectic manifold involves a three-form $\omega$ as in the case
  of geometric $H$-flux compactifications. In contrast to the
  symplectic case, in general there is no relation between
  multisymplectic and Nambu--Poisson structures. However, if $M$ has dimension
  $n+1$ then $\omega$ is a volume form on $M$ which can be inverted
  to give a Nambu--Poisson
  structure $\omega^{-1}$ on $M$. Such manifolds serve as multiphase
  spaces in {Nambu mechanics}, generalizing the usual
  Poisson phase spaces in Hamiltonian dynamics. They are the starting
  point for a formalism of \emph{higher quantization}, which we now
  describe in the context of non-geometric $Q$-flux backgrounds
  following~\cite{Saemann2011,Saemann2013}.

\subsection{Closed string noncommutative and nonassociative geometry\label{sec:NCAgeometry}}

Recall that geometric quantization of a symplectic manifold
$(M,\omega)$ is based on the requirement that the symplectic structure
defines an integer cohomology class $[\omega] \in
  \RH^2(M,\IZ)$ which encodes a {prequantum line bundle}
  with connection $(L,\nabla)$ whose first Chern class in Chern--Weil theory
  is represented by the curvature two-form $F_\nabla= 2\pi \ii \omega$. Upon prescribing a
  suitable polarization, geometric quantization amounts to
  constructing a Hilbert space of sections of the line bundle $L\to M$
  and a quantization map under which functions on $M$ act as operators
  on this Hilbert space.

In a similar fashion, a $2$-plectic manifold $(M,\varpi)$ with $[\varpi] \in \RH^3(M,\IZ)$
encodes a {prequantum abelian gerbe} with
2-connection $(\mathcal{G},A,B)$ such that $H:=\dd B =2\pi \ii \varpi$
represents the {Dixmier--Douady class} of the gerbe. It is difficult to make
sense of the notions of ``polarization'' and a ``Hilbert space of sections'' for a gerbe, so
we will instead employ the trick of~\cite{Saemann2011,Saemann2013}
which enables a proper quantization of closed string $Q$-flux
backgrounds; it involves mapping 2-plectic forms to symplectic forms by
  transgressing the gerbe $\mathcal{G}$ to a prequantum line bundle
  over the loop space
  of the configuration manifold $M$. This approach is nicely tailored
  to describing closed string noncommutative geometry, as it utilizes
  fundamental loop variables to describe non-geometry rather than
  points which are not present in $R$-flux backgrounds.

We start with the geometric background $M= \bbr^3$ or $M=T^3$ endowed with a
{2-plectic form} given by the constant $H$-flux $H= \frac16\, H_{ijk}\, \dd
x^i\wedge \dd x^j\wedge \dd x^k$. Consider the
correspondence
\begin{equation}
\xymatrix{
 & \call M\times S^1 \ar[dl]_{\rm ev} \ar[dr]^{\oint} & \\
M &  & \call M
}
\label{eq:correspondence}\end{equation}
where $\call M=C^\infty(S^1,M)$ is the loop space of $M$ parametrizing
the configuration space of
closed strings in the $H$-flux background; the map ${\rm ev}$
denotes the evaluation of a loop at a point of $S^1$ while the
map $\oint$ is integration along the $S^1$ fibre. The transgression map
$$
\calT=\mbox{$\big(\oint\big)_!$} \circ {\rm ev}^* \, :\,
\Omega^{n+1}(M) \ \longrightarrow \ \Omega^n(\call M)
$$
is defined by pulling back differential forms on $M$ via the
evaluation map of (\ref{eq:correspondence}) and then pushing forward
along the integration over $S^1$. It can be expressed locally at a loop
$x(\tau)$ as
\begin{equation*}
 (\calT\alpha)_x\big(v_1(\tau)\,,\,\ldots\,,\,v_n(\tau) \big) =
 \oint\, \dd\tau\
 \alpha\big(v_1(\tau)\,,\,\ldots\,,\,v_n(\tau)\,,\,\dot{x}(\tau) \big)
\end{equation*}
for an $n+1$-form $\alpha$ on $M$ by filling in its first $n$ slots
with parametrized vector fields $v_a(\tau)$ on $M$ and its last slot with the velocity
vector $\dot x(\tau)$ which is the natural tangent vector to the loop
$x(\tau)$.

Via the transgression map, the background $H$-flux on $M$ yields a symplectic two-form on loop space
$\call M$ given by
$$
\calh:=\calT H= \frac12\, \oint\, \dd \tau \ H_{ijk}\, \dot x^k(\tau)\, \delta
x^i(\tau)\wedge \delta x^j(\tau) \ .
$$
This two-form is always invertible and its inverse yields a Poisson
bracket on $\call M$ with
$$
\{f,g\}_{Q} := \oint \, \dd \tau\ Q^{ij}{}_{k}\,
\frac{\dot x^k(\tau)}{|\dot x(\tau)|^2}\, \Big(\, \delder{x^i(\tau)}f \,
\Big) \, \Big(\,
\delder{x^j(\tau)}g \,\Big) \ ,
$$
where the dual $Q$-flux $Q^{ij}{}_k$ is naturally identified with the
inverse of the $H$-flux $H_{ijk}$. Quantization of this Poisson
structure is rather involved and is dealt with
in~\cite{Saemann2011,Saemann2013} using an approach to quantization
based on integrating the natural Lie algebroid structure on the
Poisson manifold $\call M$ to a Lie groupoid convolution algebra. The
main virtue of this approach is that it constructs the abstract
algebra directly without concern about its representation on a Hilbert
space, which is an appropriate arena for considerations involving
nonassociativity which at the same time
avoids many of the technically
cumbersome constructions of geometric quantization. We
will not enter into any of the technical details of this quantization
procedure, which involves complicated analysis on the
infinite-dimensional loop space $\call M$,
except to note that the ensuing quantization map acts on the loop
space coordinates by sending $x^i(\tau) \mapsto \hat x^i(\tau)$ with
the commutation relations
\begin{equation*}
 \big[\hat x^i(\tau)\,,\,\hat x^j(\rho)\big] =\ii\, \hbar\, Q^{ij}{}_k\
 \widehat{\dot x^k(\tau)} \ \delta(\tau-\rho)+\calo(Q^2) \ ,
\end{equation*}
where the higher order terms are described in~\cite{Saemann2013}.
After integration over the loop parameters $\tau, \rho\in S^1$, at
leading order in the $Q$-flux this gives the commutation relations
\beq
[x^i,x^j] =\ii\hbar\,
  Q^{ij}{}_k\, w^k \quad \ , \quad [x^i,w^j]=0=[w^i,w^j]
\label{eq:closedstringNC}\eeq
where $x^i= \oint\, \dd \tau\ x^i(\tau)$ are the closed string zero
modes while $w^i = \oint\, \dd\tau\ \dot x^i(\tau)\in\IZ$ are the winding
modes. This result shows why the T-fold is only locally geometric:
Closed strings which wind acquire a position noncommutativity
proportional to the non-geometric $Q$-flux and the winding numbers. As
the relations (\ref{eq:closedstringNC}) describe a Heisenberg Lie
algebra, the resulting quantum geometry is still associative.

The same closed string noncommutativity relations were obtained by a
linearized conformal field theory analysis
in~\cite{Lust2010,Blumenhagen2011,Condeescu2012,Andriot2013}. To
linear order in the $H$-flux one can neglect the curvature
backreaction of the geometry and still work on flat target space. Then
one can quantize the closed string sigma-model in the constant
$Q$-flux background by regarding it as a left-right asymmetric
conformal field theory on a freely acting orbifold. As a closed string
winds around the base circle of the local fibration
$M\xrightarrow{T^2} S^1$ of the T-fold, the fibre coordinates
need only be periodic up to an ${\rm SL}(2,\IZ)$ automorphism of the torus
$T^2$. For the parabolic flux model, which we treat throughout this
paper, the monodromies of $T^2$ lie in a parabolic conjugacy class of
${\rm SL}(2,\IZ)$. The twisted boundary conditions on the fibre string
fields induced by the winding numbers $w^k$ and by the parabolic
${\rm SL}(2,\IZ)$ monodromies are completely analogous to those which arise
for open strings in a constant $B$-field background (see
e.g.~\cite{Chu1999}); as usual, closed strings in a twisted sector of
the orbifold conformal field theory can be regarded as open strings on
the universal covering space of the orbifold. The resulting canonical
structure and flat space mode expansion
then reproduces the relations (\ref{eq:closedstringNC}) among closed
string coordinates.

This construction can be used to obtain a quantization of the $R$-flux
background by applying formal T-duality, which sends the $Q$-flux
$Q^{ij}{}_k $ to the $R$-flux $ R^{ijk}$ and the closed string winding
variables $w^k$ to the momentum modes $ p_k$ which are canonically
conjugate to the string position coordinates $x^i$. Under this
transformation the commutation relations (\ref{eq:closedstringNC}) are
mapped to
\beq
[x^i,x^j]= \ii \hbar\, R^{ijk}\, p_k \quad , \quad
[x^i,p_j]=\ii\hbar\, \delta^i{}_j \quad , \quad
[p_i,p_j] =0 \ .
\label{eq:closedstringNA}\eeq
As noted by~\cite{Lust2011,Mylonas2012}, these relations arise from a twisted
Poisson structure on the cotangent bundle (phase space) $T^*M$; in
particular, closed string
  nonassociativity is manifested in the nonvanishing Jacobiator
\beq
[x^i,x^j,x^k]= 3\,\hbar^2\, R^{ijk}
\label{eq:closedstringJac}\eeq
exactly as it emerged in the magnetic field analog. Nonassociativity
in this instance arises from the non-trivial position-momentum
bracket, so that the antisymmetric brackets (\ref{eq:closedstringNA}) define
only a pre-Lie algebra. In the following we shall describe the
quantization of these non-geometric backgrounds to all orders in the
flux. An important ingredient will be a certain open string interpretation of
closed string correlation functions, based on the observation that
open strings do not decouple from gravity in
$R$-space~\cite{Ellwood2006} and hence open and closed strings may
become indistinguishable in these backgrounds. In~\cite{Lust2010} it
is in fact argued that closed string momentum and winding modes define
a sort of notion of D-brane in closed string theory. A path integral
description of this closed/open string duality as a result of the
asymmetric twisted sectors of the underlying worldsheet conformal
field theory is given in~\cite{Mylonas2012}.

\section{Geometrization of non-geometry}

\subsection{Generalized, doubled and phase space geometry}

Let us now describe some of the techniques proposed for
casting non-geometric spaces into local and global geometric
formulations. A common feature to all of these geometric descriptions is that
they involve some sort of a doubling of the geometry which is
precisely what makes them non-geometric: In any local description
the backgrounds require these additional geometric variables (such
as winding coordinates or momenta below). As transition functions of a
smooth manifold $M$ of dimension $d$ are valued in the structure group of the tangent
bundle, the existence of stringy ${\rm O}(d,d)$ transition functions leads
to a notion of some sort of generalized tangent bundle of rank $2d$.

One approach to the study of non-geometric fluxes is provided by
generalised geometry (see
e.g.~\cite{Grange2007,Grana2009,Halmagyi2009}). The key geometric
object in this description is the generalized tangent bundle
\beq
C=TM\oplus T^*M
\label{eq:gentanbundle}\eeq
over the target space $M$. This bundle has a natural
${\rm O}(d,d)$-invariant metric, and a generalized metric which encodes the
usual Riemannian metric and $B$-field. An abelian subgroup of suitable ${\rm O}(d,d)$
transformations, called $\beta$-transforms, of
sections of $C$ generate the non-geometric fluxes starting from the
standard geometric description in terms of a
metric, $B$-field and dilaton field, which are then expressed in terms of a bivector
field $\beta=\frac12\, \beta^{ij}\, \partial_i\wedge\partial_j$; in particular, for vanishing metric flux one has
$$
Q^{ij}{}_k=\partial_k\beta^{ij} \quad , \quad R=[\beta,\beta]_S
$$
where $[-,-]_S$ is the Schouten bracket which is the natural extension
to multivector fields of the Lie bracket of vector fields.
As this transformation is not globally defined, this
leads into a non-geometric framework. The generalized tangent bundle
(\ref{eq:gentanbundle}) will play a prominent role in this paper
within the guise of Courant algebroids.

Another approach to the study of non-geometric fluxes is provided by
doubled geometry and double field
theory~\cite{Hull2005,Hull2009,Hohm2010} (see
e.g.~\cite{Aldazabal2013,Berman2013,Hohm2013} for reviews). In this
approach one complements the spacetime coordinates $x^i$ on an equal
footing with their
dual coordinates $\tilde x_i$ which are canonically conjugate
to the winding numbers $w^i$, i.e., one first doubles the geometry $x^i \rightarrow (x^i,\tilde x_i)$ and $\partial_i \rightarrow
  (\partial_i,\tilde\partial^i)$. Doubled geometry is related to
  generalized geometry by dropping the winding coordinates while
  keeping the generalized tangent bundle (\ref{eq:gentanbundle}). The action for double field theory
  is cast in these variables using the generalized metric of
  generalized geometry and is manifestly invariant under
  ${\rm O}(d,d)$-transformations which act by rotating the doubled
  coordinates. Upon implementing the projection
  $\tilde\partial^i=0$, one recovers the usual action of
  supergravity. In particular, by performing a formal T-duality
  transformation and a field redefinition in the double field theory
  action using the ${\rm O}(d,d)$ transforms of the generalized geometry
  formalism, one can in this way obtain a field theory on $M$ for the
  non-geometric fluxes~\cite{Andriot2012} (see
  also~\cite{Blumenhagen2012}). An alternative target
  space perspective is described in~\cite{AndriotBetz2013} at purely
  the supergravity level. This theory can be reproduced from double
  field theory by dropping some degrees of freedom and it can be
  rewritten in the variables of generalized geometry. It provides a
  geometric description with non-geometric fluxes of some
  non-geometric backgrounds, demonstrating that some non-geometries
  can be described purely at the supergravity level without passing to
  more involved target space frameworks or to a worldsheet formalism.

In the following we shall review our proposal~\cite{Mylonas2012}
for the geometrization of $R$-flux backgrounds in terms of the
dynamics of a membrane sigma-model. In this approach the effective
target space seen by non-geometric strings now involves a doubling of
$M$ to its cotangent bundle $T^*M$ which is interpreted as phase space, i.e., the spacetime coordinates
$x^i$ are complemented by their conjugate momenta $p_i$. In this
setting, the $R$-flux is a geometric three-form on $T^*M$ which
represents the Dixmier--Douady class of an abelian gerbe on momentum
space. Because of the open/closed string duality of $R$-space, the
open membranes have a double life: On the one hand their dimensional reductions
are closed strings which propagate in the non-geometric flux
background, while on the other hand their boundaries (in the form of
worldvolume branch cuts) are open strings and the membrane sigma-model can be
recast as an open string $R$-twisted Poisson
sigma-model with target space $T^*M$ whose perturbative quantization induces a nonassociative
dynamical star product of fields; the corresponding Jacobiator
quantizes the closed string 3-bracket.

\subsection{$n$-algebroids and AKSZ sigma-models}

In the next section we shall develop the systematic quantization of the
nonassociative $R$-flux backgrounds by computing suitable topological sigma-model
correlation functions. A general framework to describe the sorts of
sigma-models we are interested in is provided by the AKSZ
construction~\cite{Alexandrov1997}, which builds Chern--Simons type action
functionals in the Batalin--Vilkovisky formalism for sigma-model
quantum field theories
whose target space is a symplectic Lie $n$-algebroid $E \rightarrow
M$. Recall that a Lie algebroid is a vector
bundle $E$ over a smooth manifold $M$ together with an anchor map
$$
\xymatrix{
E \ \ar[dr] \ar[rr]^{\rho} & & \ TM \ar[dl] \\
 & \ M \
}
$$
and a Lie bracket $[-,-]_E$ on sections $C^\infty(M,E)$ which is
compatible with $\rho$ and the standard Lie bracket
$[-,-]_{TM}$ between vector fields on the tangent bundle $TM$. A Lie
algebroid simultaneously generalizes the notions of Lie algebra and
tangent bundle: If $M$ is a point then a Lie algebroid is the same
thing as a Lie algebra, while $E=TM$ is always a Lie algebroid with the
identity anchor map. A more nontrivial example is provided by the case
when $M$ is a Poisson manifold with Poisson bivector $\beta$; then the
cotangent bundle $E=T^*M$ is a Lie algebroid with anchor map
$\rho=\beta^\sharp:T^*M\to TM$ defined by
$\beta^\sharp(e^i)=\beta^{ij}\, e_j$ and Lie bracket given by the Koszul bracket on
one-forms: $[e^i,e^j]_K=(\partial_k\beta^{ij})\, e^k$. Lie algebras
can always be integrated to Lie groups via the exponential map, and
conversely the tangent space of a Lie group at the identity is a Lie
algebra. Likewise, there is a corresponding notion of Lie groupoid which
is a (small) category whose object and morphism sets are smooth
manifolds, and for which every morphism is invertible. However, while
there is a suitable notion of differentiation of a Lie groupoid which
is a Lie algebroid, not every Lie algebroid can be integrated to a Lie
groupoid. A Lie $n$-algebroid is a categorification of a Lie
algebroid which is the lowest member of the hierarchy with $n=1$; we shall not give here the general definition, as in this
paper we are only interested in the cases with $n=1$ and $n=2$.

Let us begin with the case $n=1$, which is suited to the description
of strings in background $B$-fields. Then the most general
two-dimensional topological field theory that can be obtained from the
AKSZ construction is the Poisson sigma-model~\cite{Schaller1994} with target space a Lie
algebroid $E\to M$ equipped with a dual section $\Theta\in
C^\infty(M,\bigwedge^2 E^*)$. Here we will only consider the case
$E=T^*M$ with bivector field $\Theta\=\frac12\, \Theta^{ij}(x)\,
\partial_i\wedge \partial_j$. Then the action of the Poisson
sigma-model is given by
$$
S_{\rm AKSZ}^{(1)} = \int_{\Sigma_2}\, \Big(\xi_i\wedge \dd X^i+\frac12\,
\Theta^{ij}(X)\, \xi_i\wedge \xi_j \Big) \ ,
$$
where
$X: \Sigma_2 \rightarrow M$ are embedding fields of a string
worldsheet $\Sigma_2$ in the spacetime $M$ and $\xi \in
\Omega^1(\Sigma_2,X^*T^*M)$ are auxilliary one-form fields on $\Sigma_2$ with
values in the cotangent bundle $T^*M$. This field theory describes the first-order form of the topological
sector of a string sigma-model obtained by scaling the target space
metric to zero (the Seiberg--Witten limit), and in the background of a
$B$-field which is dual to the bivector $\Theta$; this can be seen by
integrating out the one-form field $\xi$ explicitly, which appears
quadratically in the action. Consistency of the equations of
motion requires $[\Theta,\Theta]_S=0$ on-shell. In this case $\Theta$ defines a
Poisson structure on $M$, with the vanishing Schouten bracket being
equivalent to the Jacobi identity for the corresponding Poisson
bracket on $C^\infty(M)$, and $E=T^*M$ is a Lie algebroid; in the
general case $E$ is only a quasi-Lie algebroid, with the trivector
$[\Theta,\Theta]_S$ controlling the violation of the Jacobi identity for
the Koszul bracket. When $\Sigma_2$ is an open Riemann surface, the
perturbative expansion of the sigma-model path integral reproduces (in
both on-shell and off-shell cases) the
Kontsevich formality maps for formal deformation
quantization~\cite{Kontsevich2003,Cattaneo2000}.

The next member of the hierarchy of AKSZ sigma-models is the case
$n=2$; a symplectic Lie 2-algebroid is the same thing as a Courant
algebroid $E\to M$. A Courant algebroid is an extension of the notion of Lie
algebroid which is further equiped with a fibre metric $h_{IJ}=
  \langle\psi_I,\psi_J\rangle$, where $\psi_I$ is a local basis of
  sections for
  $C^\infty(M,E)$. The anchor matrix $\rho(\psi_I)=
  P_I{}^i(x)\,e_i$ and the three-form
  $$
T_{IJK}(x)=[\psi_I,\psi_J,\psi_K]_E:=\mbox{$\frac16$} \, \big\langle
  [\psi_I,\psi_J]_E\,,\,\psi_K \big\rangle+ \mbox{cyclic}
$$
satisfy various conditions. The presence of
  $T_{IJK}(x)$ makes the corresponding sigma-models suited to describe
  dynamics in background three-tensor fluxes. There is a
  one-to-one correspondence between Courant algebroids and
  three-dimensional topological field theories obtained from the AKSZ construction, which are called
  Courant sigma-models~\cite{Hofman2002,Ikeda2003,Roytenberg2007} and
  are described by the action
\begin{eqnarray}
S_{\rm AKSZ}^{(2)} = \int_{\Sigma_3}\, \Big(\phi_i\wedge \dd X^i+\frac12\,
h_{IJ}\, \alpha^I\wedge \dd\alpha^J-P_I{}^i(X)\, \phi_i\wedge
\alpha^I +
\frac16\, T_{IJK}(X)\, \alpha^I\wedge\alpha^J \wedge\alpha^K \Big) \ .
\label{eq:SAKSZ2}\end{eqnarray}
Here
$X:{\Sigma_3} \rightarrow M$ are the embedding fields of a
three-dimensional membrane worldvolume $\Sigma_3$ in spacetime $M$, $\alpha\in
\Omega^1(\Sigma_3,X^*E)$ are auxilliary one-form fields on $\Sigma_3$
valued in the Courant algebroid $E$, and $\phi \in
\Omega^2({\Sigma_3},X^*T^*M)$ are auxilliary two-form fields on
$\Sigma_3$ valued in the cotangent bundle $T^*M$. In this paper we shall
work only with the standard Courant algebroid, which in
a suitable frame for the $\beta$-transformation
symmetry is the
generalized tangent bundle $E=C= TM\oplus T^*M$ with metric given by the
natural dual pairing
$$
\langle Y_1+\lambda_1, Y_2+ \lambda_2 \rangle =
\lambda_2(Y_1)+ \lambda_1(Y_2)
$$
for vector fields $Y_1,Y_2$ and one-forms $\lambda_1,\lambda_2$ on $M$,
  and the anchor map is the projection $\rho: C\rightarrow TM$. In the
  natural frame
$\psi_I= (e_i,e^i)$ the ${\rm O}(d,d)$-invariant metric has only the
non-vanishing components
$$
\langle
e_i,e^j\rangle = \delta_i{}^j \ .
$$

\subsection{Sigma-models for geometric fluxes}

Let us consider first how to apply the AKSZ sigma-model formalism to
the case of geometric flux compactifications, focusing for
definiteness on the $H$-space duality frame. The relevant
algebroid is the standard Courant algebroid
$C=TM\oplus T^*M$ twisted by a three-form $H$-flux $H=\frac16\,
  H_{ijk}(x)\, \dd x^i\wedge\dd x^j\wedge \dd x^k$; the twisting is
  accounted for by equipping $C$ with the $H$-twisted
  Courant--Dorfman bracket
\begin{eqnarray*}
[Y_1+ \lambda_1,Y_2,\lambda_2 ]_H :=[Y_1,Y_2]_{TM} +
\call_{Y_1}\lambda_2-\call_{Y_2}\lambda_1 - \mbox{$\frac12$}\, \dd
 \big(\lambda_2(Y_1)-\lambda_1(Y_2)\big)+ H(Y_1,Y_2,-) \ ,
\end{eqnarray*}
where $\call_Y$ denotes the Lie derivative along the vector field
$Y$. The only non-vanishing Lie brackets and 3-brackets evaluated on the natural
frame are given by
\beq
[e_i,e_j]_H=H_{ijk}\, e^k \qquad , \qquad
  [e_i,e_j,e_k]_H = H_{ijk} \ .
\label{eq:Hbrackets}\eeq
Let us now substitute these structure maps into the general action
(\ref{eq:SAKSZ2}), and denote $\alpha:=
  (\alpha^i,\xi_i)$ with $(\alpha^i)\in C^\infty(\Sigma_3,X^*TM)$ and
  $(\xi_i) \in C^\infty(\Sigma_3,X^*T^*M)$. In this way we arrive at
  the action for the topological membrane~\cite{Park2001}
$$
S_{\rm AKSZ}^{(2)} = \int_{\Sigma_3}\, \Big(\phi_i\wedge \dd X^i+
\alpha^i \wedge \dd\xi_i -\phi_i\wedge \alpha^i+
\frac16\, H_{ijk}(X)\, \alpha^i\wedge\alpha^j\wedge\alpha^k \Big) \ .
$$

For open membranes with boundary the worldsheet $\Sigma_2
:=\partial{\Sigma_3} \neq \varnothing$, we can add a
  boundary term $\int_{\Sigma_2} \, \frac12\,
\Theta^{ij}(X)\, \xi_i\wedge \xi_j$ for a fixed bivector $\Theta\in
C^\infty(M,\bigwedge^2TM)$ to get the action of the boundary/bulk open
topological membrane~\cite{Hofman2002}. Integrating out the two-form
fields $\phi_i$ sets $\alpha^i=\dd X^i$, and using Stokes' theorem we
arrive finally at the AKSZ string action
\begin{eqnarray*}
\widetilde{S}_{\rm AKSZ}^{\,(1)} = \int_{\Sigma_2} \, \Big(\xi_i\wedge \dd X^i+\frac12\,
\Theta^{ij}(X)\, \xi_i\wedge \xi_j \Big) + \int_{\Sigma_3} \ \frac16\,
H_{ijk}(X)\, \dd X^i\wedge\dd X^j\wedge\dd X^k \ .
\end{eqnarray*}
This is the action of the $H$-twisted Poisson sigma-model with target
space $M$~\cite{Klimcik2002}; the volume term here is completely
analogous to a Wess--Zumino--Witten term. Consistency of the equations of motion
now leads to the on-shell condition
$
[\Theta,\Theta]_S =\mbox{$\bigwedge^3$} \Theta^\sharp(H),
$
where $\Theta^\sharp(H)$ is the natural way of contracting the
three-form $H$ into a trivector using $\Theta$. This condition means
that the bivector $\Theta$ now defines an
$H$-twisted Poisson structure; in this case the Jacobi identity for
bracket $\{f,g\}_\Theta=\Theta(\dd f,\dd g)$ is violated, with the
Jacobiator controlled by the trivector $[\Theta,\Theta]_S$, i.e.,
$\{f,g,h\}_\Theta= \bigwedge^3\Theta^\sharp(H)(\dd f,\dd g,\dd
h)$. This is precisely the sigma-model that governs the topological
sector of open string dynamics in non-constant $B$-fields; in
particular, its correlation functions reproduce the nonassociative
star products of Kontsevich's deformation quantization of twisted
Poisson structures~\cite{Cornalba2002}.

\subsection{Sigma-models for non-geometric fluxes}

Now let us turn to the AKSZ sigma-model formalism appropriate
to non-geometric flux compactifications. We shall see that, in
contrast to the $H$-space sigma-models, the appropriate $R$-space
theory really is a membrane sigma-model, not a string theory, which
geometrizes the non-geometric $R$-flux background, analogously to the
way in which M-theory geometrizes the nonperturbative dynamics of
string theory. Whether or not
these membranes are fundamental degrees of freedom and related to the M2-branes of
M-theory is not clear at present. This question could be investigated
by finding a suitable topological twisting of the supersymmetric
worldvolume theory of an M2-brane. As the Courant sigma-model is the
unique three-dimensional topological field theory susceptible to the
Batalin--Vilkovisky formalism, it should then coincide with the
topological sector of this worldvolume theory. Further evidence for
this relation is provided by the fact that the noncommutative loop
algebra from \S\ref{sec:NCAgeometry} agrees with the noncommutativity
induced on the boundaries of open M2-branes ending on an M5-brane in a
constant $C$-field background~\cite{Saemann2011,Saemann2013}, which
can be interpreted as the noncommutative geometry experienced by
closed strings in constant $H$-flux backgrounds; this perspective
helps to connect the open and closed string points of view required below.

Again we start with the general Courant sigma-model (\ref{eq:SAKSZ2}) for the standard
Courant algebroid
$C= TM\oplus T^*M$, but this time twisted by a trivector flux $R=\frac16\,
R^{ijk}(x) \,\partial_i\wedge\partial_j\wedge\partial_k$. This
twisting is described by the
Roytenberg bracket
\begin{eqnarray*}
 [Y_1+ \lambda_1,Y_2+ \lambda_2 ]_R := [Y_1,Y_2]_{TM}+R(\lambda_1,\lambda_2,-)
 +
 \call_{Y_1}\lambda_2-\call_{Y_2}\lambda_1 -\, \mbox{$\frac12$}\, \dd
 \big(\lambda_2(Y_1)-\lambda_1(Y_2)
 \big) \ ,
\end{eqnarray*}
which when evaluated on the natural frame for $C$ yields the
non-vanishing Lie brackets and 3-brackets
$$
[e^i,e^j]_R=R^{ijk}\, e_k \qquad , \qquad
  [e^i,e^j,e^k]_R = R^{ijk}
$$
that are naturally dual to the bracket relations
(\ref{eq:Hbrackets}). Substituting these structure maps into the
action (\ref{eq:SAKSZ2}), and integrating out the two-form field
$\phi$ as previously yields the action
\begin{eqnarray}
{S}_R^{(2)} =\int_{\Sigma_2}\,
\xi_i\wedge\dd X^i + \int_{\Sigma_3} \ \frac16\,
R^{ijk}(X)\, \xi_i\wedge\xi_j\wedge\xi_k +
\int_{\Sigma_2} \ \frac12 \, g^{ij}(X)\, \xi_i\wedge *
\xi_j \ ,
\label{eq:Rsigmagen}\end{eqnarray}
where $g$ is a Riemannian metric on $M$ and $*$ is the Hodge duality
operator associated to a chosen metric on the worldsheet
$\Sigma_2$. Here we have explicitly broken the topological symmetry of
the sigma-model by adding a metric dependent term, in order to ensure
that a non-vanishing $R$-flux is consistent with the equations of
motion; although this may seem like a somewhat \emph{ad hoc} procedure
at the classical level, similar metric dependences would appear
anyway in the quantum action through gauge fixing terms.

The membrane sigma-model (\ref{eq:Rsigmagen}) has thus far been
written for a general $R$-flux compactification. Let us now assume
that $R^{ijk}$ and $g^{ij}$ are constant. The equation of motion for
$X$ implies that $\xi$ is a closed one-form on $\Sigma_2$, and hence
we can write $\xi_i= \dd P_i$ for some section $P\in
C^\infty(\Sigma_2,X^*T^*M)$; we do not include possible harmonic
one-form contributions to this expression, as they would drop out of
the final expressions below anyway. Using Stokes' theorem, the action
then reduces
to a pure boundary action, which we can linearize with auxiliary
fields $\eta_I$ to get a generalized Poisson sigma-model
$$
S_R^{(2)} = \int_{\Sigma_2}\, \Big(\eta_I\wedge\dd X^I+\frac12\,
\Theta^{IJ}(X) \, \eta_I\wedge\eta_J\Big) +
\int_{\Sigma_2} \ \frac12\, G^{IJ}\, \eta_I\wedge*\eta_J
$$
where
$X^I= (X^1,\dots,X^d,P_1,\dots,P_d)$ is interpreted as the embedding of the string
worldsheet $\Sigma_2$ in the cotangent bundle $T^*M$ and
\beq
\Theta=\begin{pmatrix} R^{ijk}\, p_k & \delta^i{}_j \\
  -\delta_i{}^j & 0 \end{pmatrix} \qquad , \qquad
G^{IJ} = \begin{pmatrix} g^{ij} & 0 \\ 0 &
  0 \end{pmatrix} \ .
\label{eq:Thetatwisted}\eeq
Thus the effective target space of the non-geometric string theory is
  the \emph{phase space} of the spacetime $M$.

From this sigma-model perspective, the quantity $\Theta$ is an
$H$-twisted Poisson bivector on phase space $T^*M$, with non-vanishing
Schouten bracket
$$
\Pi := [\Theta,\Theta]_S =
  \mbox{$\bigwedge^3$} \Theta^\sharp(H)
$$
where
\beq
H=\dd B \qquad , \qquad B=
  \mbox{$\frac16$} \, R^{ijk}\, p_k\, \dd p_i\wedge\dd p_j
\label{eq:U1gerbe}\eeq
is the curvature of a ${\rm U}(1)$ gerbe on momentum space. It determines a
noncommutative/nonassociative phase space with twisted Poisson
brackets
$$
\big\{x^I,x^J \big\}_\Theta=\Theta^{IJ}(x)
$$
whose quantization reproduces the closed string commutation relations
(\ref{eq:closedstringNA}). The corresponding Jacobiator
$$
\big\{x^I,x^J,x^K \big\}_\Theta := \Pi\big( x^I, x^J,
x^K \big) = \begin{pmatrix} R^{ijk} & 0\\ 0
  & 0 \end{pmatrix}
$$
quantizes the closed string 3-brackets (\ref{eq:closedstringJac}).

\section{Quantization of $\mbf R$-flux string vacua}

\subsection{Path integral quantization}

We would now like to compute correlation functions of suitable
operators in the $R$-space sigma-model. Ultimately, one would like to
do this directly at the level of the membrane sigma-model, as this is
what geometrizes the $R$-flux background. However, quantization of the
topological membrane theory is extremely difficult, as even the
gauge-fixed action is immensely complicated; part of the problem is
that, in addition to the usual gauge symmetries, the AKSZ sigma-models
in general contain higher Lie algebroid symmetries and so require
ghosts-for-ghosts in addition to the usual ghost fields. Instead, we
can exploit the hidden open string that is implicit in the open
membrane formulation of the closed string theory in $R$-space. The
multivaluedness of the closed string fields discussed in \S\ref{sec:NCAgeometry} are
implemented in the underlying orbifold conformal field theory via
insertions of twist fields in correlators which create branch cuts on the
worldsheet. By extending the closed string worldsheet to a membrane
worldvolume $\Sigma_3$, the resulting branched surface can be
interpreted as an open string worldsheet; see~\cite{Mylonas2012} for
further details of the path integral description of this closed/open
string duality. An alternative perspective would be to allow for
singularities and assume that the closed string worldsheet
is a surface with corners; in that case one could in principle deal
with the boundary correlation functions that we discuss below, but
such an approach seems technically cumbersome and we do not know how
to proceed with this point of view.

From this open string perspective, suitable functional integrals in
the generalized Poisson sigma-model then reproduce Kontsevich's graphical expansion for global
  deformation quantization~\cite{Cattaneo2000}. The key quantities
  that emerge from these perturbative computations are the Kontsevich formality
  maps $U_n$, which take $n$ multivector fields ${\cal
    X}_1, \dots ,{\cal X}_n$ on
  $\CM:=T^*M$ to a multidifferential operator
$$
U_{n}({\cal X}_1, \dots ,{\cal X}_n) = \sum_{\Gamma\in G_n}\, w_\Gamma\,
D_\Gamma({\cal X}_1,\dots,{\cal X}_n) \ ,
$$
where the sum is taken over all admissible diagrams $\Gamma$ and the
weights $w_\Gamma$ of graphs are computed from certain integrals over
geodesic angles in the hyperbolic upper half-plane regarded
as the disk $\Sigma_2$. The geodesics represent derivatives emanating from the
multivector fields which act on functions that are inserted on the
boundary $\partial\Sigma_2\cong \bbr$ of the upper half-plane. For
example, the action of the bivector $\Theta= \frac12\,
\Theta^{IJ}\, \partial_I\wedge \partial_J$ is represented as
$$
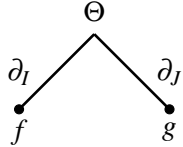
\begin{tikzpicture}[scale=1.0,>=stealth, transform shape]
\draw [line width=1] (3,1.5)--(2,0.5);
\draw [line width=1] (4,0.5)--(3,1.5);
\draw [fill=black] (2,0.5) circle (.6mm);
\draw [fill=black] (4,0.5) circle (.6mm);
\node at (2,1) {$\partial_I$};
\node at (4,1) {$\partial_J$};
\node at (2,0.2) {$f$};
\node at (4,0.2) {$g$};
\node at (3,1.8) {$\Theta$};
\end{tikzpicture}
$$
and it computes the star product
\begin{eqnarray*}
f\star g = \sum_{n=0}^{\infty}\, {\frac{(\ii\hbar)^n}{n!}\,
   U_{n}(\Theta, \dots ,\Theta )} (f,g) =: \Phi(\Theta)(f,g) \ ,
\end{eqnarray*}
while the action of the trivector $\Pi=\frac16\,
\Pi^{IJK}\, \partial_I\wedge \partial_J\wedge \partial_K=[\Theta,\Theta]_S$ is depicted by
$$
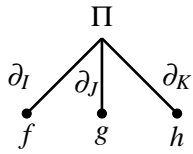
\begin{tikzpicture}[scale=1.0,>=stealth, transform shape]
\draw [line width=1] (8,0.5)--(9,1.5);
\draw [line width=1] (9,0.5)--(9,1.5);
\draw [line width=1] (10,0.5)--(9,1.5);
\draw [fill=black] (8,0.5) circle (.6mm);
\draw [fill=black] (9,0.5) circle (.6mm);
\draw [fill=black] (10,0.5) circle (.6mm);
\node at (7.9,1) {$\partial_I$};
\node at (8.8,0.9) {$\partial_J$};
\node at (10,1) {$\partial_K$};
\node at (8,0.2) {$f$};
\node at (9,0.2) {$g$};
\node at (10,0.2) {$h$};
\node at (9,1.8) {$\Pi$};
\end{tikzpicture}
$$
and it computes a 3-bracket
\begin{eqnarray*}
[f,g,h]_\star = \sum_{n=0}^\infty\, {\frac{(\ii\hbar)^n}{n!}\,
  U_{n+1}(\Pi,\Theta,\dots,\Theta)}(f,g,h) =: \Phi(\Pi)(f,g,h) \ ,
\end{eqnarray*}
where $f,g,h\in C^\infty(\CM)$.

The maps $U_n$ define
$L_\infty$-morphisms of differential graded Lie algebras relating
Schouten brackets $[-,-]_S$ to
  Gerstenhaber brackets $[-,-]_G$, which are the natural extensions to
  multidifferential operators of the
  commutator bracket of differential operators;
  in particular, they satisfy formality
conditions~\cite{Kontsevich2003}. For example, the formality condition $[\Phi(\Theta),\star]_G= \ii\hbar\,\Phi([\Theta,\Theta]_S)$ explicitly
  quantifies nonassociativity of the star product through the
  3-bracket as
$$
(f\star g)\star h - f\star (g\star h) = \mbox{$\frac{\hbar}{2\ii}$}
\, \Phi(\Pi)(f,g,h) = \mbox{$\frac{\hbar}{2\ii}$} \, [f,g,h]_\star  \ .
$$
The formality conditions also imply derivation properties. For
instance, they map Hamiltonian vector fields to inner derivations of
the star product and quasi-Poisson vector fields to differential
operators which are derivations of the star
product~\cite{Mylonas2012}. On the other hand,
the formality condition $[\Phi(\Pi),\star]_G\= \ii\hbar\,\Phi([\Pi,\Theta]_S)$
  encodes a quantum analogue of the Leibniz rule for the {Nambu--Poisson
    structure} $\{f,g,h\}_\Pi :=\ \Pi(\dd f,\dd g,\dd h)$, which for
  the case at hand with constant $R$-flux implies
\beq
[ f \star g , h, k]_\star -   [f,  g\star h,
 k]_\star + [f, g, h\star k]_\star = f \star [g,h,k]_\star
 +[f,g,h]_\star \star k \ .
\label{eq:derivprop}\eeq
Thus our construction also
provides a means for quantizing Nambu--Poisson structures, which is a
notoriously difficult unsolved problem in general (see
e.g.~\cite{DeBellis2010} and references therein).

Explicit formulas can be obtained from the fact that all Kontsevich
diagrams factorize and their weights can be expressed in terms of
three diagrams (up to permutations), two involving the bivector field
$\Theta$ and one involving the trivector field
$\Pi$~\cite{Mylonas2012}. This yields the dynamical
\emph{nonassociative} star-product
\beq
f\star g = f\star_p g  := {\mbf\cdot}\, \big(\e^{\frac{\ii\hbar}{2}\, R^{ijk}\, p_k \,
  \partial_i \otimes \partial_j}\, \e^{\frac{\ii \hbar}2\, (\partial_i \otimes \tilde\partial^i -
  \tilde\partial^i \otimes \partial_i)}(f \otimes g ) \big) \ ,
\label{eq:fstarpg}\eeq
where $\partial_i=\frac\partial{\partial x^i}$ and
$\tilde\partial^i=\frac\partial{\partial p_i}$. Nonassociativity
arises when the derivatives $\tilde\partial^i$ hit explicit momenta
$p_i$ in
the bidifferential operator defining the star product. By replacing
the dynamical
momentum variable $p$ with a constant momentum $\bar p$, we obtain an
associative Moyal--Weyl type star-product
  $\bar\star := \star_{\bar p}$. We can then express triple products
in
  terms of a tridifferential operator as
\beq
(f\star g)\star h = \Big[ \,\bar\star\, \Big(
\exp\big(\mbox{$\frac{\hbar^2}{4}$} \, R^{ijk} \, \partial_i \otimes\partial_j
\otimes \partial_k \big)(f
\otimes g \otimes h)\Big)\,
\Big]_{\bar p \to
  p} \ ,
\label{eq:star3product}\eeq
where no ordering is required on the right-hand side because of
associativity of the star product $\bar\star$, and the notation $\bar
p\to p$ denotes the reinstatement of dynamical momentum. Nonassociativity is expressed through the explicit formula for the 3-bracket given by
$$
[f,g,h]_\star =\frac{4\ii}\hbar\, \Big[ \,\bar\star\, \Big(
\sinh\big(\mbox{$\frac{\hbar^2}{4}$} \, R^{ijk} \, \partial_i \otimes\partial_j
\otimes \partial_k \big)(f
\otimes g \otimes h)\Big)\,
\Big]_{\bar p \to
  p} \ .
$$

This star
product reproduces the fundamental phase space commutation relations
\beq
[x^i\stackrel{\star}{,}x^j]= \ii \hbar\, R^{ijk}\, p_k \qquad , \qquad
[x^i\stackrel{\star}{,}p_j]=\ii\hbar\, \delta^i{}_j \qquad , \qquad
[p_i\stackrel{\star}{,}p_j] =0 \ ,
\label{eq:fundcommrels}\eeq
and it possesses the desired physical properties anticipated from
on-shell closed string scattering amplitudes. For example, both
2-cyclicity 
and 3-cyclicity hold, i.e.,
\bea
\int_{{\cal M}}\, {\dd^{2d}x \ f \star g} &=&\int_{{\cal M}}\,
{\dd^{2d}x \ g \star f}  \=\int_{{\cal M}}\,
{\dd^{2d}x \ f\, g}\ , \nonumber\\[4pt]
\int_{{\cal M}}\, {\dd^{2d}x \ f \star (g \star h)} &=& \int_{{\cal
    M}}\, {\dd^{2d}x \ (f \star g) \star h} \ ,
\label{2-3cyclicity}
\eea
showing
  that physical closed string states do not see noncommutativity or
  nonassociativity of the non-geometric flux background. In
  particular, when restricted to functions on configuration space $M$,
  the 3-product (\ref{eq:star3product}) reproduces the triproduct of~\cite{Blumenhagen2011}
  which was conjectured to reproduce off-shell correlation functions
  of closed string tachyon vertex
  operators in $R$-space in a linearized conformal field theory
  analysis. The 2-cyclicity and 3-cyclicity properties are also the
  basis for a consistent formulation of nonassociative phase space
  quantum mechanics.

\subsection{Seiberg--Witten maps and noncommutative gerbes}

In open string noncommutative gauge theory, the Seiberg--Witten map is
an equivalence of (associative) star products $\star$ and
$\star'$ generated by a covariantizing map $\calD$ which is a quantum analogue of Moser's lemma in symplectic
geometry.
Let $\Theta$ be a Poisson bivector, i.e., $[\Theta,\Theta]_S=0$, with
dual two-form $B=\Theta^{-1}$, and
let $\call\to \CM$ be a line bundle with curvature
$F=\dd A$. Let $\rho$ be the flow generated by the vector field
$\Theta(A,-)$. Then the map
$$
B \ \xrightarrow{ \ \rho \ } \ B+F
$$
is generated by a change of coordinates and quantizes to a map which sits in a commutative diagram~\cite{Jurco2000,Jurco2000a,Jurco2001,Jurco2002}
$$
\xymatrix{\Theta \ \ar[d]_\rho \ar[rr]^{\rm quantization} & & \ \star
  \ar[d]^\calD \\ \Theta' \ \ar[rr]_{\rm quantization} & & \ \star'}
$$
where
$\Theta'=\Theta\, (1+\hbar\,F\, \Theta)^{-1}$ is a new Poisson
bivector, and $\calD(f\star' g)=\calD f\star \calD
  g$; the noncommutative gauge field $\hat A$ is defined by the
  covariant coordinates $\calD x =: x+\hat A$ such that an ordinary
  gauge orbit of $A$ corresponds to a noncommutative gauge orbit of
  $\hat A$.

This construction can be used to realise the quantization of twisted
Poisson structures in terms of noncommutative
gerbes in the sense of~\cite{Aschieri2010}. Let $\Theta$ be a Poisson structure
twisted by a closed three-form $H$,
i.e., $[\Theta,\Theta]_S=\bigwedge^3\Theta^\sharp(H)$. Let
$\{U_\alpha\}$ be a good open covering of the manifold $\CM$. Then
on $U_\alpha$ the $H$-flux is generated by a locally defined $B$-field as
$H=\dd B_\alpha$, $B_\alpha\in\Omega^2(U_\alpha)$. On non-empty overlaps $U_\alpha\cap U_\beta$ the
difference of potentials $B_\beta-B_\alpha$ is a closed two-form,
hence $B_\beta-B_\alpha= F_{\alpha\beta}= \dd A_{\alpha\beta}$ for local one-forms
$A_{\alpha\beta}$ which determine ${\rm U}(1)$ gauge fields on a line bundle
$\call_{\alpha\beta}\to U_\alpha\cap U_\beta$. The local $B$-fields can be used to locally
untwist $\Theta$ to Poisson bivectors $\Theta_\alpha := \Theta\,
(1-\hbar\, B_\alpha\, \Theta)^{-1}$ on each patch $U_\alpha$. Then
while the quantization of $\Theta$ yields a nonassociative star
product $\star$, quantization of
$\Theta_\alpha$ yields a family of local associative star products
$\star_\alpha$ such that $\star_\alpha$ and $\star_\beta$ are
equivalent by covariantizing maps $\calD_{\alpha\beta}$ constructed
from the curvatures $F_{\alpha\beta}$ on each non-empty overlap
$U_\alpha\cap U_\beta$.

Now let us apply these general considerations to the construction of
generalized Seiberg--Witten maps for non-geometric
fluxes~\cite{Mylonas2012}. The $R$-twisted Poisson structure on
phase space $\CM=T^*M$ is generated by the ${\rm U}(1)$ gerbe on
momentum space with
curvature (\ref{eq:U1gerbe}); this is a trivial (but
not flat) gerbe,
so we can foliate $\CM$ by surfaces of constant momentum and
replace the open patch label $\alpha$ with the constant
momentum vector $\bar p$. Then the relevant two-tensors are given by
\beq
\Theta_{\bar p} = \begin{pmatrix} \hbar \, R^{ijk} \, \bar p_k &
  \delta^i{}_j \\ -\delta_i{}^j & 0 \end{pmatrix} \qquad , \qquad
B_{\bar p} = \begin{pmatrix} 0 & 0\\ 0 & R^{ijk}\, (p_k - \bar
  p_k) \end{pmatrix}
\label{eq:Thetabarp}\eeq
where $\Theta_{\bar p}$ are Poisson bivectors which untwist the
twisted Poisson bivector (\ref{eq:Thetatwisted}), and $H=\dd B_{\bar
  p}= \frac16\, R^{ijk}\, \dd p_i\wedge \dd p_j\wedge \dd p_k$ is the
curvature (\ref{eq:U1gerbe}) of the trivial gerbe. The equivalence maps
$\calD_{\bar p\bar p'}$ between associative star products $\bar\star $ and
$\bar\star'$ are generated by the gauge fields $A_{\bar p\bar p'}\= R^{ijk}\,
p_i\, (\bar p_k-\bar p_k^{\,\prime})\, \dd p_j$ with curvature
$F_{\bar p\bar p'}= \frac12\, R^{ijk}\, (\bar p_k-\bar p_k')\, \dd
p_i\wedge\dd p_j$.
In particular, for $\bar
p=0$ we recover the canonical Moyal--Weyl star-product $\star_0$ on
phase space. The map $\calD_{\bar p}$ generated by $A_{\bar p}=
R^{ijk}\, p_i\, \bar p_k\, \dd p_j$ from associative to nonassociative
star products can be computed explicitly~\cite{Mylonas2012} and satisfies
$$
f\star g= \big[\mathcal D_{\bar p} f \star_0 \mathcal D_{\bar p}
g\big]_{\bar p \rightarrow p} \ .
$$

There is moreover an explicit nonassociative generalization of the Seiberg--Witten
map suitable for the non-geometric backgrounds. A construction based
directly on a twisted Poisson bivector $\Theta$ is usually spoiled by
spurious terms involving the non-vanishing Schouten brackets
$[\Theta,\Theta]_S$. However, in the present case such difficulties
can be avoided by restricting to maps involving only gauge fields
$A=
  \bar a^i(x,p)\, \dd p_i$ which have no components along the
  configuration space $M$. There are two particular cases of interest
  in the context of this paper. Firstly, there are general coordinate
  transformations which are generated by the vector field $\Theta(A,-)= \bar
  a^i(x,p)\, \partial_i$ and are mapped to quantized diffeomorphisms;
  this lends some credibility to the hope that there is some notion of
  nonassociative gravity in non-geometric string
  backgrounds. Secondly, there are the Nambu--Poisson maps which are
  generated by $A= R(a_2,-)$ for an arbitrary two-form
  $a_2\in\Omega^2(M)$, whose quantization leads to a higher gauge
  theory of quantized Nambu--Poisson tensor fields.

\subsection{3-cocycles and categorified Weyl quantization}

Let us now briefly explain some algebraic meanings behind the
quantization of the twisted Poisson structure as alternative but
equivalent quantizations to that obtained via the more analytical
sigma-model approach. This can be elegantly
understood through the origin of the generalized Poisson sigma-model
as an AKSZ topological field theory with target space a Courant
algebroid; in this way quantization can be cast as the problem of convolution
    quantization of Lie 2-algebras, which are categorifications of Lie
    algebras.

The reduction of the
  standard $R$-space Courant algebroid from before to the case where
  the base space $M$ is a point yields a quadratic Lie algebra $\frh$ with
  commutation relations
$$
[ x^i, x^j]_Q=R^{ijk}\, {\bar p}_k \qquad , \qquad [ x^i,{\bar p}_j]_Q=0= [{\bar p}_i ,{\bar
  p}_j ]_Q
$$
and an $\frh$-invariant ${\rm O}(d,d)$-symmetric inner product with non-vanishing values
$$
\langle x^i,\bar p_j\rangle = \delta^i{}_j \ .
$$
These are just the commutation relations of a Heisenberg Lie algebra,
and they mimick the closed string relations (\ref{eq:closedstringNC})
in the $Q$-space duality frame; in particular, the Lie algebra $\frh$
can be regarded as a quantization of the Poisson structure
$\Theta_{\bar p}$ from (\ref{eq:Thetabarp}). This quadratic Lie
algebra induces in the standard way a skeletal 2-term $L_\infty$-algebra
$$
V_1 = \bbr \ \xrightarrow{\
   0 \ } \ V_0 = \frh
$$
with classifying 3-cocycle $j:\frh\wedge\frh\wedge\frh \rightarrow
\bbr$, in the Chevalley--Eilenberg cohomology of $\frh$ with values in
the trivial representation, whose sole non-vanishing values are given by
$$
j(x^i,x^j,x^k)= \mbox{$\frac16$}\, \big\langle [x^i,x^j]_Q\,,\,
x^k\big\rangle +\mbox{cyclic} = R^{ijk} \ .
$$
This Lie 2-algebra can be integrated in the usual way to a Lie 2-group
$$
\xymatrix@C=10mm{
\CCG_1=G_\frh\times {\rm U}(1) \ \ar@< 2pt>[r] \ar@< -2pt>[r] & \
\CCG_0= G
}
$$
with $G_\frh$ the Heisenberg group integrating $\frh$ via the exponential
map, whose associator integrates $j$ and defines a 3-cocycle
in the Chevalley--Eilenberg cohomology of $G_\frh$ with values in
$\bbr$. The roles of 3-cocycles of Lie algebra and Lie group
cohomology in the quantization of nonassociative $R$-space is elucidated in~\cite{Bakas2013}. By computing convolution type products
in this Lie 2-group, one can mimick the standard approach based on
Weyl quantization in the associative setting of operator algebras (see
e.g.~\cite{Szabo2003} for a review) to rederive the
nonassociative star products (\ref{eq:fstarpg}) for the phase space description of
$R$-space; this algebraic approach is taken in~\cite{Mylonas2012,Bakas2013}.

\subsection{Quasi-Hopf cochain twist quantization}

There is yet another equivalent algebraic approach to the quantization of the
$R$-flux background which emphasises the symmetries underlying the
non-geometric flux compactification, and which provides a systematic
means to obtain nonassociative deformations of geometry and gravity in
$R$-space. This approach further illustrates in what sense
nonassociative deformations really do grow naturally in the wild.

For this, let us first recall the associative setting of
Hopf cocycle twist quantization~\cite{Majid1995}. A {Drinfel'd twist}
for a Hopf algebra $H= H(\Delta,S,\varepsilon,\cdot)$ is an invertible
counital element
$$
F=F_{(1)}\otimes F_{(2)} \ \in \ H\otimes H
$$
which satisfies the 2-cocycle condition
$$
(F\otimes 1)\,\Delta_1 F= (1\otimes F)\, \Delta_2 F
$$
with $\Delta_1=\Delta\otimes1$ and $\Delta_2=1\otimes\Delta$. The
twist $F$ can be used to map $H$ to a new Hopf algebra
  $H_F= H_F(\Delta_F,S_F,\varepsilon,\cdot)$ with the same underlying
  algebraic structure but modified coalgebra structure; in particular,
  the new coproduct is given by
$$
\Delta_F= F\, \Delta\, F^{-1} \ .
$$
This construction quantizes any $H$-module algebra $A$ to a
  ``braided-commutative'' algebra $A_F$ with deformed product
$$
f \star g= {\mbf\cdot}\, \big(F^{-1}(f\otimes g)\big)= F^{-1}_{(1)}
f\, \cdot \, F^{-1}_{(2)}g
$$
for $f,g\in A$; this deformation ensures that the action of the twisted Hopf algebra
$H_F$ is compatible with the product on $A_F$, i.e., that $A_F$ is an
$H_F$-module algebra.

Now let us relax the 2-cocycle condition and consider an arbitrary
2-cochain twist $F\in H\otimes H$. In that case $H_F$ is only a
  \emph{quasi-Hopf algebra}, i.e., the coassociativity of the
  coproduct is violated in a controlled way
$$
(\Delta_F)_2\, \Delta_F = {\phi} \,(\Delta_F)_1\, \Delta_F \, {\phi^{-1}}
$$
by means of a multiplicative \emph{associator} $\phi= \phi_{(1)}\otimes
  \phi_{(2)}\otimes \phi_{(3)} \in H\otimes H\otimes H$ which is a
  3-cocycle obtained as the coboundary of the cochain twist via
$$
\phi= \partial^*F  := F_{23}\, \Delta_2\, F\, \Delta_1\, F^{-1}\,
F_{12}^{-1}
$$
where $F_{23}=1\otimes F_{(1)}\otimes F_{(2)}$ and
$F_{12}=F_{(1)}\otimes F_{(2)}\otimes 1$. The notion of quasi-Hopf
algebra was introduced in the early days of quantum groups by
Drinfel'd when it was realised that many examples of quantum universal
enveloping algebras satisfy only this weaker criterion. Now an $H$-module algebra
$A$ gets quantized to a ``quasi-associative'' $H_F$-module algebra
$A_F$ with
$$
(f\star g)\star h \= \phi_{(1)} f\star(\phi_{(2)}g\star \phi_{(3)}h) \ .
$$
The power of this approach is its generality: Any algebraic entity can
be quantized in this way, once a Hopf module structure and 2-cochain
are chosen.
For our applications we are interested in the specific example where $H= U(\frg)$ is the
enveloping Hopf algebra of a Lie algebra $\frg$ of symmetries acting on a manifold
  $\CM$; then the algebra of functions $A=C^\infty(\CM)$ can be
  quantized in this way to a generically noncommutative and
  nonassociative algebra
  $A_F$.
Similarly, the exterior algebra $\Omega^\bullet(\CM)$ of
differential forms on $\CM$ is quantized to
$\Omega^\bullet_F(\CM)$, and so on for other geometrical structures.

This construction can be formalised into the notion of a twist
quantization functor, i.e., it simultaneously
deforms all $H$-covariant constructions as a
  functorial isomorphism
$$
\calq_F\,:\, {}^H\calm~ \longrightarrow~ {}^{H_F}\calm
$$
of braided monoidal categories of left
  $H$-modules ${}^H\calm$ and left $H_F$-modules ${}^{H_F}\calm$. In the
  monoidal category ${}^{H_F}\calm$ the associator $\phi\in H\otimes H\otimes
  H$ induces non-trivial associativity isomorphisms
  $\Phi_{V,W,Z}: (V\otimes W)\otimes Z \rightarrow
V\otimes(W\otimes Z)$ defined by
$$
\Phi_{V,W,Z}\big((v\otimes w)\otimes z\big) = \phi_{(1)} v\otimes
(\phi_{(2)} w \otimes
\phi_{(3)} z) \ .
$$
The five-term 3-cocycle condition on $\phi$ implies that $\Phi$ obeys
MacLane's pentagon relations; from the perspective of quantization of
3-brackets,
the pentagon relations yield derivation properties such as
(\ref{eq:derivprop}). On the other hand, a braiding of the category
${}^{H_F}\calm$ is provided by a quasi-triangular structure on the
Hopf algebra and it induces non-trivial commutativity isomorphisms
  $\Psi_{V,W} : V\otimes W \rightarrow W\otimes V$ given by
$$
\Psi_{V,W}(v\otimes w) = F^{-2}_{(1)}w\otimes F^{-2}_{(2)} v \ .
$$
From this perspective, both noncommutativity and nonassociativity are
very natural features provided we work in the ``right'' category: While the
algebra $A_F$ is noncommutative and nonassociative when normally
considered as an object of the category of vector spaces, which has
trivial braiding and associator, it \emph{is} commutative and
associative in the category ${}^{H_F}\calm$.

Let us now apply these general considerations to obtain the cochain
twist quantization of $R$-space~\cite{Mylonas2013}. For this, let
$\frg$ be the nonabelian
  Lie algebra of phase space translations and Bopp shifts whose action
  on $C^\infty(\CM)$ is generated by the vector fields
$$
P_i= \partial_i \qquad , \qquad \tilde P{}^i= \tilde \partial{}^i \qquad , \qquad
M_{ij}= p_i\, \partial_j-p_j\, \partial_i \ .
$$
For $\sigma^{ij}=-\sigma^{ji}\in\bbr$, the vector fields
$\sigma^{ij}\,M_{ij}$ leave momenta fixed and act on position
coordinates as the familiar non-local Bopp shifts
$x^i \mapsto x^i+\sigma^{ij}\, p_j$ from quantum mechanics which mix
positions with momenta. They define a quasi-Hopf
deformation of the universal enveloping algebra $U(\frg)$ by the
cochain twist
$$
\CF= \exp\left[ \mbox{$-\frac{\ii\hbar}{2}$}\, \left(\mbox{$\frac
        14$}\,R^{ijk}\, \big(M_{ij}\otimes P_k - P_i \otimes M_{jk}
      \big) + P_i \otimes \tilde P^i - \tilde
      P^i \otimes P_i\right)\right] \ ,
$$
and the quantization functor on the category of quasi-Hopf module algebras
generates nonassociative algebras through the associator
$$
\phi=\partial^*\CF= \exp\big(\mbox{$\frac{\hbar^2}{2}$} \,R^{ijk}\, P_i \otimes P_j
\otimes P_k \big) \ .
$$
The corresponding deformed product on the algebra $C^\infty(\CM)$
coincides with the star product (\ref{eq:fstarpg}), while the associator
reproduces the triple product structure (\ref{eq:star3product}) and
coincides with the 3-cocycle encountered before in a different context.

To illustrate the utility of this approach in constructing
deformations of geometry, let us work out the corresponding
nonassociative exterior differential calculus. We start with the usual
exterior algebra complex $(\Omega^\bullet(\CM), \wedge, \dd)$ of
differential forms on phase
space, and assume that the exterior derivative $\dd$ is equivariant
under the covariant
  action of the enveloping Hopf algebra $H=U(\frg) $. Then the action
  of $H$ on $\Omega^\bullet(\CM)$ is given by Lie
  derivatives $\CL_h$ along the vector fields corresponding to
  elements $h\in H$; on differentials the only non-trivial actions are given by
$$
M_{ij} \, \dd x^k := \CL_{M_{ij}}(\dd x^k) = \delta_j{}^k \,\dd p_i
-\delta_i{}^k \, \dd p_j \ .
$$
The deformed exterior product $\omega\wedge_\star \eta:=
  \wedge\big(F^{-1}(\omega\otimes \eta)\big)$ is generically noncommutative and
  nonassociative with the basic relations
$$
\dd x^I \wedge_\star \dd x^J = - \dd x^J \wedge_\star \dd x^I = \dd
x^I \wedge \dd x^J \ ,
$$
$$
\big( \dd x^I \wedge_\star \dd x^J\big)\wedge_\star \dd x^K = \dd x^I
\wedge_\star \big(\dd x^J\wedge_\star \dd x^K \big) \ .
$$
By the equivariance condition, the exterior derivative $\dd$ does not
undergo any deformation as it is a morphism of the category
${}^{H_F}\calm$. The action of $C^\infty(\CM)$ on $\Omega^\bullet(\CM)$ by pointwise
multiplication of a form with a function is quantized to a deformed
$A_\CF$-bimodule structure whose only non-trivial relations are given by
$$
x^i \star \dd x^j =\dd x^j \star x^i +\mbox{$\frac{\ii
  \hbar}{2}$}\,R^{ijk}\,\dd p_k \ .
$$
The nonassociative differential calculus also obeys the requisite physical
properties of graded 2-cyclicity
$$
\int_\CM \, {\omega \wedge_\star \eta}
=(-1)^{|\omega|\, |\eta|}\, \int_\CM\,
{\eta \wedge_\star \omega} = \int_\CM \, {\omega \wedge \eta}
$$
and graded 3-cyclicity
$$
\int_\CM\, {(\omega \wedge_\star \eta) \wedge_\star \lambda} =
\int_\CM \, {\omega \wedge_\star (\eta \wedge_\star
  \lambda)} \ .
$$

\subsection{Quantum mechanics with a 3-cocycle}

The commutation relations (\ref{eq:closedstringNA}) capture the
nonassociative geometry that arises in $R$-flux backgrounds and are
heuristically expected to lead to novel uncertainty principles for
spacetime positions. However, a rigorous derivation requires a
formulation of quantum mechanics adapted to a nonassociative
setting. This excludes all standard associative operator algebra
approaches and is complicated further by the fact that our algebras
are not of Jordan type, which is essentially the only case for which
nonassociative aspects of quantum mechanics have been studied so
far. A generalization of the phase space formulation of quantum
mechanics, based on the nonassociative star product (\ref{eq:fstarpg})
with fundamental commutation relations (\ref{eq:fundcommrels}),
appears to be the most convenient approach. An essential ingredient for consistency
are the 2-cyclicity and \mbox{3-cyclicity} properties
(\ref{2-3cyclicity}) of the star product with respect to integration.
In the following we give a brief overview of nonassociative quantum
mechanics and some of its consequences, refering to~\cite{Mylonas2013}
for further details.

An observable $A$ in this approach is a real-valued function on
$2d$-dimensional phase space $\CM$. More generally, operators are
complex-valued functions and are multiplied together with the star
product (\ref{eq:fstarpg}). Dynamics is implemented via
Heisenberg-type time evolution equations
\beqa
\frac{\partial A}{\partial t} = \frac \ii\hbar \,
[H\stackrel{\star}{,} A]
\eeqa
that look familiar, but in general are not derivations of the
nonassociative star product operator algebra; similar evolution equations allow the study of the motion of a charge particle in a magnetic field with sources, i.e., $\nabla\cdot \mbf B \neq 0$.
States are characterized by normalized (complex-valued) functions
$\psi_\alpha$ and statistical probabilities $\lambda_\alpha$ (for
mixed states). Expectation values are computed via the phase space integral
\beqa
\langle A \rangle = \sum_{\alpha=1}^n \, \lambda_\alpha \ \intps \psi_\alpha^* \star (A \star \psi_\alpha)  = \intps A \, S \ ,
\eeqa
and can be expressed as indicated in terms of a normalized real-valued
state function 
$$
S = \sum_{\alpha=1}^n \, \lambda_\alpha
\, \psi_\alpha \star \psi_\alpha^*
$$
 using (\ref{2-3cyclicity}). Given a state, we can define a semi-definite sesquilinear form for operators
\beqa
(A,B) 
= \sum_{\alpha=1}^n \, \lambda_\alpha \ \intps (A \star \psi_\alpha)^* \, (B \star \psi_\alpha)
\eeqa
that satisfies the Cauchy-Schwarz inequality 
\beq \label{Cauchy-Schwarz}
\big|(A,B)\big|^2 \leq  (A,A)\, (B,B) \ .
\eeq

The inequality (\ref{Cauchy-Schwarz}) is the basis for the derivation of uncertainty relations 
\[
\Delta p_i \, \Delta p_j \geq 0 \qquad , \qquad
\Delta x^i \, \Delta p_j \geq \mbox{$\frac\hbar 2$} \, \delta^i{}_j \qquad , \qquad
\Delta x^i \, \Delta x^j \geq \mbox{$\frac\hbar 2$}\, \big| R^{ijk} \,
\langle p_k \rangle\big| \ ,
\]
and encodes the concept of positivity of operators in our setting. 
Eigen-state functions $S$ of operators $A$ with eigenvalues $\lambda
\in \mathbb C$  are defined as usual, i.e., $A \star S = \lambda S$,
and  observables ($A^* = A$) have real eigenvalues. Due to
nonassociativity, this fact is not straightforward, as $(A \star S)
\star A^* \neq A \star (S \star A^*)$; the proof requires 3-cyclicity (\ref{2-3cyclicity}).
As in ordinary quantum mechanics, a pair of operators that do not
commute cannot in general be measured simultaneously to arbitrary
precision. A famous example are the limitations to position and
momentum measurements imposed by the Heisenberg uncertainty
principle. Only commuting operators have complete sets of common
eigenstates. In nonassociative quantum mechanics, similar statements
hold for triples of operators that do not associate. In particular any
triple of coordinates $x^i$, $x^j$, $x^k$ that do not associate,
i.e., with non-trivial $R$-flux $R^{ijk} \neq 0$, do not have any
common eigenstate. The $R$-flux induced nonassociativity thus leads to
a coarse-graining of spacetime with fundamental limitations to the determination of the exact locations of events.
These new uncertainties are quantified by the non-zero expectation
values of area and volume operators given by~\cite{Mylonas2013}
\[
\langle A^{ij} \rangle = \hbar \, R^{ijk} \,\langle p_k\rangle
\qquad , \qquad \langle V^{ijk} \rangle = \mbox{$\frac32$} \, \hbar^2\, R^{ijk} \ .
\]

\section*{Acknowledgments}

R.J.S. thanks the organisors of the satellite conference for the
opportunity to present these lectures and the warm hospitality during
the meeting, and in particular Andrey Bytsenko for the invitation. The work of
D.M. is supported by
the Greek National Scholarship Foundation. The work of D.M. and R.J.S. was supported in part by the Consolidated Grant ST/J000310/1
from the U.K. Science and Technology Facilities Council. The work of P.S. was supported by the DFG RTG 1620 ``Models of Gravity''.


\end{document}